\documentclass[preprint,nofootinbib,aps,superscriptaddress,eqsecnum]{revtex4-1} 
\pdfoutput=1
\usepackage{extarrows}
\usepackage{mathtools}
\usepackage{siunitx}
\usepackage{graphicx}
\usepackage{amsmath}
\usepackage{amsfonts}
\usepackage{amssymb}
\usepackage{hyperref}
\usepackage{caption}
\usepackage{subcaption}
\usepackage{color}
\allowdisplaybreaks
\def\bea{\begin{eqnarray}}
\def\eea{\end{eqnarray}}
\def\be{\begin{equation}}
\def\ee{\end{equation}}

\begin{document}

	\title{Dark matter and low scale leptogenesis in a flavor symmetric neutrino two Higgs doublet model($\nu$2HDM)}
	
		\author{Lavina Sarma}
	\email{lavina@tezu.ernet.in}
	\author{Bichitra Bijay Boruah}
	\email{bijay@tezu.ernet.in}
\author{Mrinal Kumar Das}
	\email{mkdas@tezu.ernet.in}
	\affiliation{Department of Physics, Tezpur University, Tezpur 784028, India}

	\begin{abstract}
		We have studied dark matter (DM) phenomenology, neutrinoless double beta decay (NDBD) and realized low scale leptogenesis in an extension of Standard Model(SM) with three neutral fermions, a scalar doublet and a dark sector incorporating a singlet scalar and a Dirac singlet fermion. A generic model based on $A_4\times Z_4$ flavor symmetry including five flavons is used to explain both normal and inverted hierarchy mass patterns of neutrino and also to accommodate the dark matter mass. In this extension of the $\nu$2HDM, the effective neutrino mass observed in $0\nu\beta\beta$ is well within the experimental limit provided by KamLAND-ZEN. In order to validate DM within this model, we have checked relic abundance and free streaming length of the dark sector component, i.e. a Dirac singlet fermion constraining its mass in keV range. More importantly we have also realised low scale leptogenesis simultaneously within this framework and also the Dirac CP phase gets constrained with the results. Bound from LFV is also incorporated in order to constrain the Yukawa couplings. More importantly, we have analyzed the dependence of various phenomenology with decay parameter for different choice of arbitrary complex angles.
	\end{abstract}
	
	\maketitle	
	
	\section{Introduction}
	
	The significance of physics Beyond the Standard Model(BSM) can be mainly verified by 
	the existence of neutrino mass \cite{deSalas:2017kay}, dark datter(DM)\cite{bertone2005particle,Moore:1999nt} and baryon asymmetry of the Universe(BAU)\cite{leptogenesis,Hugle:2018qbw}. The mass of neutrinos to be of the order of sub-eV \cite{Lattanzi:2016rre} got its confirmation from the neutrino oscillations\cite{neuOsc,neuOsc2}. Neutrino oscillation was first suggested by Pontecorvo in the 1960's. Whereas, its experimental discovery was made by Super-Kamiokande Observatory and Sudbury Neutrino Observatories which was recognised by the 2015 Noble Prize shared by Takaaki Kajita and Arthur B. McDonald. Neutrino physics has come up with some benchmark evidences for developing new physics for elementary particles thereby probing into the evolution of the Universe. The recent Neutrino experiments MINOS\cite{MINOS},RENO\cite{PhysRevLett.108.191802},T2K\cite{T2K},Double-Chooz\cite{DCHOOZ} have not only confirmed but also measured the neutrino oscillation parameters more accurately. In case of Type I seesaw \cite{Minkowski,Mohapatra}, the Standard Model is extended with the help of three right handed neutrinos which results in the generation of neutrino masses at sub eV scale whereas the right handed neutrino(RHN) mass is related to the Grand Unified Theory(GUT) scale. Though the heavy RHN can generate leptogenesis\cite{Fukugita} and simulataneously the small neutrino mass, a naturalness problem arises\cite{naturalness}. Some possible pathways to overcome the high scale leptogenesis are through resonant leptogenesis\cite{ResLEP}, ARS(Akhmedov, Rubakov and	Smirnov) proposed a mechanism via neutrino oscillation wherein the baryon asymmetry of the Universe
	could be generated with the help of CP-violating sterile neutrino oscillations\cite{ARS,ARS1}, or via Higgs decay\cite{HiggsDECAY,HiggsDECAY1}. Such mechanisms however demand for a degenerate RHN mass, which further is another naturalness problem. Whereas, we do have models like the $\nu$2HDM\cite{nu2HDM} and scotogenic model\cite{Scotogenic,Lavina,Borah:2018rca} which accounts an explanation for these discrepencies by considering hierarchical masses of RHN and thereby producing leptogenesis at TeV scale. In this work, we have taken the $\nu$2HDM and studied various phenomenologies persisting in it.

	There have been numerous lines of evidence of DM including observations from galaxy rotation curve\cite{Rubin:1970zza}, cosmic microwave background\cite{cosmicmicrowave}, galaxy cluster\cite{Zwicky:1933gu}, gravitational lensing\cite{Treu:2012sn} , etc. However, the most recent cosmological constraints on DM comes from the Planck satellite\cite{Ade:2015fva}. It ascertains 27$\%$ of the present Universe to constitue of DM, which accounts for five times the abundance of the baryonic matter. Depending upon the internal structure i.e. the constituent particles and symmetries, the history of thermalisation of the dark sector take different paths. This gives rise to different types of DM on the basis of its mass range\cite{FIMP}. We basically have the FIMP, hot and cold type dark matter followed from the above mentioned aspect. Though the lightest of the RHN can be considered to be a sterile DM in keV scale\cite{N1DM,N1DM1}, it faces stringent bounds from various experiments\cite{Sterile}, thereby leaving a very small viable region for it to be a probable dark matter candidate. So, we introduce another scenario in our work by incorporating a dark sector to the $\nu$2HDM. It includes a scalar singlet($\eta$) and a Dirac singlet fermion($\xi$), which are charged under $Z_{2}$ symmetry\cite{darksector}. Interestingly, the bounds from X-ray experiment can be ignored as the stability of DM candidate $\xi$ is maintained by the $Z_{2}$ symmetry\cite{Sterile}. Due to the fact that we have null results from the Direct detection\cite{PandaX} as well as indirect detection\cite{Fermi-LAT}, the nature of the DM candidate is taken to be of FIMP type.
	
		The basic idea followed in our work comprises of the realisation of the $\nu$2HDM by the virtue of the flavor symmetries $A_{4}\otimes Z_{4}$. Along with the constituent particles of the generic $\nu$2HDM, we have added a singlet scalar and a Dirac singlet fermion to the model. The newly added particles incorporate the dark sector, by which the DM phenomenology can be explained. As already mentioned, the DM considered in our work is of FIMP type. 
	We have analysed leptogenesis in our work with the consideration of a mass hierarchy of the RHN $M_{N_{1}}<< M_{N_{2,3}}$ . On finding the numerical values of the Yukawa coupling matrix, we further carry out a detailed analysis of leptogenesis as well as neutrinoless double beta decay($0\nu\beta\beta$). Also the branching ratio of the decay $(\mu \rightarrow e\gamma)$ is calculated in order to constrain the Yukawa couplings obtained from the model. We emphasize on a relative study due to the variation in the values of the rotational matrix angles used for parametrization of the Yukawa coupling matrix. This yields variation in decay parameter value $K_{N_{1}}$, which is used in determining baryon asymmetry of the Universe and also explains the various DM phenomenologies. Thus, we see how the choice of the angles has an impact on neutrino phenomenology and related cosmology carried out in this work.
	
	We have further categorised the paper into six sections which are as follows. Sec.\eqref{sec3} includes the basic model and the flavor symmetric model. Phenomenologies such as leptogenesis and dark matter are mentioned in sec.\eqref{sec4} and sec.\eqref{sec5} respectively. We finally show the numerical analysis and the results obtained in sec.\eqref{sec6}, followed by the conclusion in sec.\eqref{sec7}.
	
	   \section{ Flavor symmetric neutrino two higgs doublet model($\nu$2HDM)}\label{sec3}
	    In spite of being the most successful theory of particle physics, Standard model(SM) fails to explain quite a few experimental and theoritical phenomena such as smallness of neutrino mass, dark matter(DM), baryon asymmetry of the Universe(BAU), etc. In the SM, neutrino remains massless, although it can accquire a small Majorana mass through dimension five operator\cite{Weinberg1,Ma11}.
	    \begin{equation}
	    \frac{1}{\Lambda}(\nu_{i} \Phi^{0}-l_{i} \Phi^{+})(\nu_{j} \Phi^{0}-l_{j} \Phi^{+})
	    \end{equation}\label{o}
	    where $\Lambda$ is the effective large mass scale and H is usual Higss doublet. This operator can be commonly realized in the framework of canonical seesaw\cite{Mohapatra}, where SM is extended with three right handed neutrino($N_{i}$) such that-
	    \begin{equation} \label{e}  
	    m_{\nu}=\frac{m_{D}^{2}}{m_{N}}
	    \end{equation}
	    where $m_{D}=fv$ is the Dirac neutrino mass considering vaccum expection value(vev) of Higgs to be $<\Phi^{0}>=v'$ and $m_{N}=\Lambda f$. Such kind of mechanisms does not have a direct test in the experiment beacuse of the large mass scale of $m_{N}$.
	    
	    $\rm\nu2HDM$ is one such framework where Eq\eqref{e} can be realised naturally with mass of the right handed neutrino to be in experimental reach (order of 1 TeV) by extending the SM with three right handed neutrino and a scalar doublet($\phi$) having a very small vaccum expectation value (vev)\cite{nu2HDM}. We have to introduce a $U(1)_{L}$ global symmetry under which $L_{\Phi}=0$, $L_{\phi}=-1$ and  $L_{N}=0$, so that it forbids type-I seesaw interaction term $\bar{l_{L}}\tilde{\Phi}N$. Thus, the additional scalar doublet $\phi$ couples with right handed neutrino and SM Higgs will couple with quarks and charged lepton. In this model the smallness of Dirac neutrino mass is achieved through this additional scalar doublet. To accomodate dark matter candidate in this model, we introduce a dark sector to the generic $\nu$2HDM(discussed above).This newly added dark sector includes a singlet scalar $\eta$ and a Dirac singlet fermion $\xi$\cite{lowscale,FIMP} which are charged under $Z_{2}$ symmetry. Considering $m_{\xi}< m_{\eta}$, we can say that $\xi$ serves as a probable DM candidate. Particle content of the minimal $\rm\nu2HDM$ with charge assignment under $SU(3)_{C} \otimes SU(2)_{L} \otimes U(1)_{Y}$ are given below-
	    
	    \begin{equation}\label{eqx}
	    \left[\begin{array}{cc}
	    \nu_{i}\\
	    l_{i}
	    \end{array}\right]_L=(1,2,-1/2) ;  \left[\begin{array}{c}
	    l_{R_{i}}
	    \end{array}\right] =(1,1,-1), \left[\begin{array}{c}
	    N_{i}
	    \end{array}\right] =(1,1,0),  
	    \left[\begin{array}{c}
	    \phi
	    \end{array}\right] =(1,2,1/2) .
	    \end{equation}

	    
	   The scalar doublets $\Phi$ and $\phi$ in the model can be denoted as:

	  \begin{equation}
	    \Phi=\left[\begin{array}{cc}
	    \Phi^{+}\\
	    \frac{v'+\Phi^{0,r}+i\Phi^{0,i}}{\sqrt{2}}
	    \end{array}\right]\\,
	    \phi=\left[\begin{array}{cc}
	    \phi^{+}\\
	    \frac{v+\phi^{0,r}+i\phi^{0,i}}{\sqrt{2}}
	    \end{array}\right]
	    \end{equation}
	   
	   Thus, the corresponding Higgs potential will be:
	    \begin{equation}
	    \begin{split}
	    V= &m_{\Phi}^{2}\Phi^{\dagger}\Phi+m_{\phi}^{2}\phi^{\dagger}\phi+ m_{\eta}^{2}\eta^{\dagger}\eta+ \frac{1}{2}\lambda_{1}(\Phi^{\dagger}\Phi)^{2}+\frac{1}{2} \lambda_{2}(\phi^{\dagger}\phi)+    \lambda_{3}(\Phi^{\dagger}\Phi)(\phi^{\dagger}\phi)+\\&\lambda_{4}(\Phi^{\dagger}\phi)(\phi^{\dagger}\Phi)-\mu_{12}^{2}(\Phi^{\dagger}\phi+h.c)+(\frac{1}{2} \lambda_{5}(\Phi^{\dagger}\phi)^{2}+h.c)+\frac{1}{2} \lambda_{6}(\eta^{\dagger}\eta)^{2}+\lambda_{7}(\eta^{\dagger}\eta)(\Phi^{\dagger}\Phi)+\lambda_{8}(\eta^{\dagger}\eta)(\phi^{\dagger}\phi)
	    \end{split}
	    \end{equation}
	    where the lepton symmetry is explicitly and softly broken by the $\mu_{12}$
	    term. The condition $<\eta>=0$ should be satisfied for the unbroken $Z_{2}$ symmetry so that it donot have any interaction with the Higgs doublet. Also the mixing angle tan$\beta$ can be expressed interms of the vev of SM Higgs boson and the inert Higgs doublet, which is given by
	    $tan\beta= \frac{v^{'}}{v}$.\\
	      Again the minimization condition is given by the equation:
	   \begin{equation}
	     m_{\Phi}^{2}= \mu_{12}^{2}\frac{v}{v^{'}} - \frac{\lambda_{1}}{2}v'^{2} -\frac{(\lambda_{3}+\lambda_{4}+\lambda_{5})}{2}v^{2}
	    \end{equation}                                           
	    	    \begin{equation}
	     m_{\phi}^{2}= \mu_{12}^{2}\frac{v^{'}}{v} - \frac{\lambda_{1}}{2}v^{2} -\frac{(\lambda_{3}+\lambda_{4}+\lambda_{5})}{2}v'^{2}
	   \end{equation}     
	    which helps in expressing the vev of Higgs doublet in terms of parameter present in the Higgs potential.
	    Radiative corrections to the term $\mu_{12}^{2}$ is proportional to $\mu_{12}^{2}$ itself and it is logarithimically sensitive to the cut off scale as it is the only source of lepton number violation\cite{LeptonV}. This results in the stabilization of vev hierarchy $v<<v^{'}$ against radiative corrections. The physical Higgs boson after SSB is given by:
	    \begin{equation}
	    	H^{+}= \phi^{+}cos\beta - \Phi^{+}sin\beta, A= \phi^{0,i}cos\beta - \Phi^{0,i}sin\beta
	    \end{equation}
	    \begin{equation}
	    H^0= \phi^{0,r}cos\alpha - \Phi^{0,r}sin\alpha, h= \Phi^{0,r}cos\alpha + \Phi^{0,r}sin\alpha,
	    \end{equation}
	    where the mixing angle $\alpha$ is given as follows:
	  \begin{equation}
	    	tan 2\alpha \simeq 2\frac{v}{v^{'}}\frac{-\mu_{12}^{2}+(\lambda_{3}+\lambda_{4}+\lambda_{5})v^{'}v}{-\mu_{12}^{2}+\lambda_{1}v^{'}v}.
	    \end{equation}
    Further after neglecting $\mathcal{O}(v^{2})$ and $\mathcal{O}(\mu_{12}^{2})$, the masses of the physical Higgs bosons are as follows:
\begin{equation}
m_{H^{+}}^{2}\simeq m_{\phi}^{2}+1/2(\lambda_{3}v^{'2}), m_{A}^{2}\simeq m_{H}^{2}\simeq m_{H^{+}}^{2}+1/2(\lambda_{4}+\lambda_{5})v^{'2}, m_{h}\simeq \lambda_{1}v^{'2}
\end{equation}
	    The effective Lagrangian for the extended $\nu$2HDM can be expressed as:
	    \begin{equation}
	    \mathcal{L} \subset Y \bar{l_{L}} \tilde{\phi}N + \lambda\bar{\xi}\eta N + \frac{1}{2} \bar{N^{c}}m_{N} N+ m_{\xi} \xi \bar{\xi}+h.c \label{L1}.
	    \end{equation}
	    From eq\eqref{L1}, the light neutrino mass can derived as\cite{Minkowski}:
	    \begin{equation}
	    m_{\nu}=-\frac{v^{2}}{2} Y M^{-1}_{N} Y^{T}= U_{PMNS} m^{d}_{\nu} U^{T}_{PMNS}\label{mnu} 	
	        \end{equation}
	    where, $m^{d}_{\nu}=diag(m_{1},m_{2},m_{3})$ is the diagonal neutrino mass matrix and $U_{PMNS}$ is the PMNS(Pontecorvo-Maki-Nakagawa-Sakata) matrix.

	 We also realize the extension of the $\nu$2HDM with the help of discrete flavor symmetry group $A_{4}\otimes Z_{4}$. To maintain the stability of the dark sector, a discrete $Z_{2}$ symmetry is also introduced under which the extended particles are charged. Particle content of the model with its charge assignment are given in the table\eqref{TAB}. $\eta$ and $\xi$ fields are charged under $Z_{2}$ symmetry, considering them to be the dark sector which donot acquire any vev and thus remain invisible. We introduce five flavon fields $\rho$, $\rho^{'}$, $\rho^{''}$, $\zeta$, $\zeta^{'}$ in order to break the flavor symmetry so as to generate the required mass structures. 
	       \begin{table}[h]
	    	\begin{center}
	    		\begin{tabular}{|c|c|c|c|c|c|c|c|c|c|c|c|c|c|c|c|c|c|}
	    			
	    			\hline 
	    			Field &$l_{L}$ & $l_{R_{1}}$ & $l_{R_{2}}$ & $l_{R_{3}}$ & $N_{1}$ & $N_{2}$& $N_{3}$ &$\Phi$& $\phi$ & $\rho$ &$\rho^{'}$&$\rho^{''}$& $\zeta$&$\zeta^{'}$ \\ 
	    			\hline
	    			$SU(2)$&$2$&$1$&$1$&$1$&$1$&$1$&$1$&$2$&$2$&$1$&$1$&$1$&$1$&$1$\\
	    			\hline
	    			$A_{4}$& $3$ & $1$ & $1^{''}$ &$1^{'}$& $1$ & $1^{'}$ & $1$ &  $1$ &$1$&$3$&$3$&$3$&$1$&$1^{'}$ \\
	    			\hline 
	    			$Z_{4}$& $1$ & $1$ &$1$& $1$ & $1$ & $-i$ &  $-1$ &$1$&$1$&$1$&$i$&$-1$&$1$&$-1$\\
	    			\hline 
	    		\end{tabular} 
	    	\end{center}
	    	\caption{Fields and their respective transformations
	    		under the symmetry group of the model.} \label{TAB}
	    \end{table}

    Now, the Lagrangian for the charged lepton sector can be written as:
    
    \begin{equation}
    \mathcal{L}= \frac{y_{e}}{\Lambda} (\bar{l_{L}}\Phi\rho)l_{R_{1}}+\frac{y_{\mu}}{\Lambda} (\bar{l_{L}}\Phi\rho)l_{R_{2}}+\frac{y_{\tau}}{\Lambda} (\bar{l_{L}}\Phi\rho)l_{R_{3}}+h.c \label{i2}
    \end{equation}
    
    where $\Lambda$ denotes the cut off scale. The term $\bar{l_L}H\rho$ in the Eq.\eqref{i2} transforms as $1$, $1^{'}$ and $1^{''}$ respectively under $A_{4}$ for initial three terms in order to obtain a diagonal charged lepton mass matrix\cite{Zhang}.
    
	 The effective Lagrangain for Dirac mass is given by:
		\begin{equation}
	\mathcal{L}= \frac{y_{1}}{\Lambda} (\bar{l_{L}}\tilde{\phi}\rho) N_{1}+\frac{y_{2}}{\Lambda} (\bar{l_{L}}\tilde{\phi}\rho^{'}) N_{2}+\frac{y_{3}}{\Lambda} (\bar{l_{L}}\tilde{\phi}\rho^{''}) N_{3}
	\end{equation} 
	and the right handed fermions are represented by the Lagrangian:
	\begin{equation}
		\frac{1}{2}\beta_{1}\zeta \bar{N_{1}^{c}}N_{1}+\frac{1}{2}\beta_{2}\zeta^{'} \bar{N_{2}^{c}}N_{2}+\frac{1}{2}\beta_{3}\zeta \bar{N_{3}^{c}}N_{3}+h.c
		\end{equation} 
	The flavon alignments considered in the model are given below:
	\begin{align}
	<\rho>=(\omega,\omega,\omega),  <\rho^{'}>=(0,-\omega,\omega),  <\rho^{''}>=(\omega,\omega,\omega), <\zeta>=<\zeta^{'}>=\omega
	\end{align}
Using this alignment we can have diagonal charge lepton matrix given by:
 
\begin{equation}
m_{l}=\frac{<\Phi> \omega}{\Lambda}\left[\begin{array}{ccc}
y_{e}&0&0\\
0&y_{\mu}&0\\
0&0&y_{\tau}
\end{array}\right]\\,
\end{equation}

and the Dirac mass matrix will take the form represented as:
\begin{equation}
m_{D}=\frac{<\phi> \omega}{\Lambda}\left[\begin{array}{ccc}
y_{1}&y_{2}&y_{3}\\
y_{1}&0&y_{3}\\
y_{1}&-y_{2}&y_{3}
\end{array}\right]\\,
\end{equation}
here $y_{1}$,$y_{2}$ and $y_{3}$ stands for the Yukawa couplings. We further express the Dirc mass matrix in terms of the model parameters as:

\begin{equation}
m_{D}=\left[\begin{array}{ccc}
a&b&c\\
a&0&c\\
a&-b&c
\end{array}\right]\\,
\end{equation}
where, $a=\frac{<\phi>\omega }{\Lambda}y_{1}$, $b=\frac{<\phi>\omega }{\Lambda}y_{2}$ and $c=\frac{<\phi>\omega }{\Lambda}y_{3}$ are the model parameters which actually represents the Dirac masses. 
Because of the additional $Z_{4}$ symmetry the right handed neutrino mass matrix will be a diagonal one given as:

\begin{equation}
M_{N}=\left[\begin{array}{ccc}
\beta_{1}\omega&0&0\\
0&\beta_{2}\omega&0\\
0&0&\beta_{3}\omega
\end{array}\right]\\,
\end{equation}
where $\beta_{1}$,$\beta_{2}$ and $\beta_{3}$ are the Majorana couplings. Incorporating the mass matrices and Yukawa coupling matrix obtained from the model in Eq.\ref{mnu}, we can finally calculate the active neutrino mass.

	 \textbf{Bounds from Lepton flavor violating process}:
	  
	  It is well known that lepton flavor violating processes put significant bound on the model parameter space. The size of the LFV is controlled by the lepton number violating Yukawa couplings $Y_{ij}$. The MEG collaboration has been able to set the impressive bound on muon decay  Br$(\mu \rightarrow e\gamma)<4.2 \times 10^{-13}$\cite{TheMEG}.
	  The branching ratio of $l_{\alpha}\rightarrow l_{\beta}\gamma$ is given by\cite{branch}-
	 \begin{equation}
	  	\text{Br}(\mu\rightarrow e\gamma)=\frac{3\alpha}{64\pi G^{2}}\bigg|\sum_{i} \frac{Y_{\mu i}Y_{ei}^{*}}{m^{2}_{H^{+}}} F(\Delta^{N_{i}}_{H^{+}})\bigg|
	  	\end{equation}
	  where G is the Fermi constant and $\Delta^{N_{i}}_{H^{+}}= (\frac{M_{N_{i}}}{m_{H^{+}}})^{2}$ and the loop function is expressed as :
	  \begin{equation}
	  F(x)= \frac{1}{6(1-x)^{4}}(1-6x^{2}+3x^{2}+2x^{3}-6x^{2}lnx)
	  \end{equation}
  The Yukawa couplings obtained from the model, i.e. found to be in the range $10^{-1}-10^{-6}$ are consistent with the bounds of lepton flavor violating process. From Fig.\eqref{B1}, we see that the branching ratio value for the process $\mu\rightarrow e\gamma$ is below the experimental bound given by MEG collaboration for both NH and IH, thereby, making the Yukawa couplings viable in explaining leptogenesis and dark matter.

				\section{Leptogenesis}\label{sec4}
				
				We study leptogenesis in this work, which is a consequence of the out-of-equilibrium decay of $ N_{1}\rightarrow  l_{L}\phi^{*},\bar{ l_{L}}\phi$ \cite{leptogenesis}. The lepton asymmetry produced only by the decay of $N_{1}$ is converted into the baryon asymmetry of the Universe(BAU) by the electro-weak sphaleron phase transitions\cite{B-L}. In case of vanilla leptogenesis, there exists an absolute lower bound on the mass of the lightest RHN to be $M_{N_{1}} \simeq 10^{9}$ GeV \cite{Davidson:2002qv,Buchmuller:2002rq}. However, the limit on the lightest RHN mass scale can be lowered to 10 TeV \cite{Hugle:2018qbw,Borah:2018rca} in certain scenarios. With the consideration of a hierarchical mass spectrum $ m_{\phi}<< M_{N_{1}}<< M_{N_{2,3}}$ and the incorporation of the Yukawa couplings from the model, the CP-asymmetry term is given by\cite{Borah:2018rca}:
				\begin{equation}
				\epsilon_{1}= \frac{1}{8\pi( Y^{\dagger} Y)_{11}}\sum_{j\ne 1} Im[( Y^{\dagger} Y)^{2}]_{1j}\left[ f( r_{j1},\eta_{1})- \frac{\sqrt{ r_{j1}}}{ r_{j1}-1}(1-\eta_{1})^{2}\right],
				\end{equation}
				where, Y represents the Yukawa coupling matrix and the term $f( r_{j1},\eta_{1})$ is expressed as: 
				\begin{equation}
				 f( r_{j1},\eta_{1})= \sqrt{ r_{j1}}\left[1+ \frac{(1-2\eta_{1}+ r_{j1})}{(1-\eta_{1})^{2}}   ln(\frac{ r_{j1}-\eta_{1}^{2}}{1-2\eta_{1}+ r_{j1}})\right],
				\end{equation}
			
				with $ r_{j1}= \big(\frac{ M_{N_{j}}}{ M_{N_{1}}}\big)^{2}$ and $\eta_{1}\equiv \big(\frac{ m_{H^{0}}}{ M_{N_{1}}}\big)^{2}$ . \\
				
				The decay rate equation for $ N_{1}$ which is given by,
				\begin{equation}
				\Gamma_{1}= \frac{ M_{N_{1}}}{8\pi}( Y^{\dagger} Y)_{11}\left[1- \Big(\frac{ m_{H^{0}}}{ M_{N_{1}}}\Big)^{2}\right]^{2}= \frac{ M_{N_{1}}}{8\pi}( Y^{\dagger} Y)_{11}(1-\eta_{1})^{2}
				\end{equation}
					Meanwhile, the washout effect is checked by the decay parameter
					\begin{equation}
						K_{N_{1}}= \frac{\Gamma_{1}}{ H(z=1)},
					\end{equation}
				where, $\Gamma_{1}$ is the decay width of $ N_{1}$, H is the Hubble parameter and $z=M_{N_{1}}/T$ with T being the temperature of the thermal bath.  We can express $ H$ in terms of $T$ and the corresponding equation is given by:
				\begin{equation}\label{eq:2}
				H = \sqrt\frac{8\pi^{3}g_{*}}{90}\dfrac{T^{2}}{M_{Pl}}.
				\end{equation} 
				In Eq.\eqref{eq:2}, $g_{*}$ stands for the effective number of relativistic degrees of freedom and $M_{Pl}\simeq 1.22\times 10^{19}$ GeV is the Planck mass. \\
				Using the parametrization for Yukawa coupling, i.e.\cite{Ibarra:2016dlb}
				\begin{equation}
				Y= \frac{\sqrt{2}}{v}U_{PMNS}(m^{diag}_{\nu})^{1/2}R(M^{diag}_{N})^{1/2}, \label{yuk}
				\end{equation} 
				we can verify the following relation:
				\begin{equation}
					Y^{\dagger}Y= \frac{2}{v^{2}}(M^{diag}_{N})^{1/2}R^{\dagger}(m^{diag}_{\nu})R(M^{diag}_{N})^{1/2}\label{11}.
				\end{equation}
				Here, R is an orthogonal matrix satisying the condition $R^{T}R$=1 which is parametrized as:
				\begin{equation}
				R=	\begin{pmatrix}
					cos\omega_{12}  & -sin\omega_{12}  & 0 \\
					sin\omega_{12}  & cos\omega_{12}  & 0\\
					0 & 0 & 1\\
					\end{pmatrix}
					\begin{pmatrix}
					cos\omega_{13}  & 0  & -sin\omega_{13} \\
				0  & 1  & 0\\
					sin\omega_{13} & 0 & cos\omega_{13}\\
					\end{pmatrix}
					\begin{pmatrix}
					1  & 0  & 0 \\
					0  & cos\omega_{23}  & -sin\omega_{23}\\
					0 & sin\omega_{23} & cos\omega_{23}\\
					\end{pmatrix}
				\end{equation}
				where, $\omega_{12,13,23}$ represents the arbitrary complex angles.
				Thus, we can now express the decay parameter $ K_{N_{1}}$ interms of the arbitrary angles by the following equation:
				\begin{equation}
					 K_{N_{1}}\simeq 897 \big(tan\beta\big)^{2}\frac{|((m_{\nu}^{diag})^{R})_{11}|}{eV}\label{3},
				\end{equation} 
				which can also be verified from Eq.\eqref{11}. 
				In the above equation, $(m_{\nu}^{diag})^{R}= R^{\dagger}m_{\nu}^{diag}R$, which further gives:
				\begin{equation}
					(m_{\nu}^{diag})^{R})_{11}= m_{1}cos\omega_{12}^{2}cos\omega_{13}^{2}+ m_{2}sin\omega_{12}^{2}cos\omega_{13}^{2}+m_{3}sin\omega_{13}^{2}\label{2}
				\end{equation} 
				From Eq\eqref{3} and Eq\eqref{2}, we can see that $K_{N_{1}}$ has dependency on the arbitrary complex angles $\omega_{12,13}$ but is independent of $\omega_{23}$. This relationship implies that by proper variation of these arbitrary angles, we can have a comparative analysis of the different phenomenologies associated with it. In our work, we have choosen some benchmark values for $\omega_{13,23}$ for which we have varied $\omega_{12}$. We showcase the variations in BAU as well as DM phenomenology for two different values of $\omega_{12}$, however, keeping $\omega_{13,23}$ fixed. Also, the vev of the scalar doublet $\phi$ plays a vital role in the determination of $K_{N_{1}}$, which further has significant impact on baryogenesis. As mentioned in various literatures\cite{lepto,lepto1}, the $\Delta L$= 0 washout processes are crucial in context of small values $u$. Thereby, the condition $\Gamma_{1}/M_{N_{1}} <<1$ for low scale seesaw is satisfied. 
				Thus, the Boltzmann equations for the evolution of the abundance of $N_{1}$ and $N_{B-L}$ are given by \cite{Davidson:2002qv},
				\begin{equation}\label{eq:3}
				\frac{dn_{N_{1}}}{dz}= -D_{1}(n_{N_{1}} - n_{N_{1}}^{eq}),
				\end{equation}
				\begin{equation}\label{eq:4}
				\frac{dn_{B-L}}{dz}= -\epsilon_{1}D_{1}(n_{N_{1}} - n_{N_{1}}^{eq})- W_{1}n_{B-L},
				\end{equation}
				respectively. $n_{N_{1}}^{eq}= \frac{z^{2}}{2}K_{1}(z)$ is the equilibrium number density of $N_{1}$, where $K_{i}(z)$ is the modified  Bessel function of $i^{th}$ type and
				\begin{equation}
				D_{1}\equiv \frac{\Gamma_{1}}{Hz} = K_{N_{1}}z\frac{K_{1}(z)}{K_{2}(z)}
				\end{equation} 
				gives the measure of the total decay rate with respect to the Hubble rate, and $ W_{1}= \frac{\Gamma_{W}}{Hz}$ is the total washout rate. Again,  $W_{1}= W_{1D}+W_{\Delta L=2}$ \cite{Hugle:2018qbw}, i.e the total washout term is the sum of the washout due to inverse decays $l\phi,\bar{l}\phi^{*}\rightarrow N_{1}$ ($W_{1D}= \frac{1}{4}K_{N_{1}}z^{3}K_{1}(z)$) and the washout due to the $\Delta L= 2$ scatterings $l\phi \leftrightarrow \bar{l}\phi^{*},ll \leftrightarrow \phi^{*}\phi^{*}$ which is given by\cite{Borah:2018rca},
				\begin{equation}
				W_{\Delta L=2} \simeq \dfrac{18\sqrt{10}M_{Pl}}{\pi^{4}g_{l}\sqrt{g_{*}}z^{2}v^{4}}(\frac{2\pi^{2}}{\lambda_{5}})^{2}M_{N_{1}}\bar{m}^{2}.
				\end{equation}
				Here, $g_{l}$ is the internal degrees of freedom for the SM leptons, and $\bar{m}$ is the absolute neutrino mass scale, defined by:
				\begin{equation}
				\bar{m}^{2} = m_{1}^{2}+m_{2}^{2}+m_{3}^{2}
				\end{equation}

				The final B-L asymmetry $ n_{B-L}^{f}$ is evaluated by numerically calculating Eq.\eqref{eq:3} and Eq.\eqref{eq:4} before the sphaleron freeze-out. This is converted into the baryon-to-photon ratio given by\cite{Borah:2018rca}:
				\begin{equation}\label{eq:5}
				n_{B}= \frac{3}{4}\frac{g_{*}^{0}}{g_{*}}a_{sph}n_{B-L}^{f}\simeq 9.2\times 10^{-3}n_{B-L}^{f},
				\end{equation} 
				In Eq.\eqref{eq:5}, $g_{*}= 110.75$ is the effective relativistic degrees of freedom at the time when final lepton asymmetry was produced, $g_{*}^{0}= \frac{43}{11}$ is the effective degrees of freedom at the recombination epoch and $ a_{sph}=\frac{8}{23}$ is the sphaleron conversion factor taking two Higgs doublet into consideration. The Planck limit 2018 gives a bound on the observed BAU($ n_{B}$) to be $(6.04\pm0.08)\times 10^{-10}$ \cite{Aghanim:2018eyx}. Therefore, in our work, we have chosen the free parameters such that we can generate the observed BAU for NH/IH. As mentioned earlier, we have fixed the arbitrary angles $\omega_{13,23}$ bearing values $10^{-3}+10^{-3}i$ and $10^{-1}+10^{-1}i$ respectively. Whereas, we have chosen two benchmark values for $\omega_{12}$, i.e $10^{-2}+10^{-2}i$ and $10^{-12}+10^{-12}i$ which shows quite a significant change in the cosmological phenomenologies carried out in this work. In addition to this, we have taken a certain parameter space for $v= 0.1-30$ GeV to show its dependence on BAU for both NH/IH. 
				\section{DARK MATTER}\label{sec5}
				In this work, we consider a FIMP type dark matter with the real Yukawa coupling, $\lambda<<1$\cite{FIMP} between the right handed neutrino N with the dark sector, i.e. singlet fermion $\xi$ and singlet scalar $\eta$. As of the DM type taken into account, it is known that the coupling is so weak that it never reaches thermalization even within the dark sector. We consider the singlet fermion($\xi$) to be a probable dark matter candidate. Due to the feeble interaction of the DM candidate, its relic abundance is generated via the freeze-in mechanism\cite{FIMP2}. We can obtain the expression for relic abundance by solving the Boltzmann equation:
				\begin{equation}
					\frac{dY_{\xi}}{dz}= DY_{N_{1}}\rm BR_{\xi}\label{4}
				\end{equation}
				where, $Y_{\xi}$ and $Y_{N_{1}}$ are the abundance of DM candidate $\xi$ and right handed neutrino $N_{1}$ respectively, and $\rm BR_{\xi}$ is the branching ratio of $N_{1}\longrightarrow \xi \eta$. The inverse decay process, i.e. $\xi \eta\longrightarrow N_{1}$ is neglected and a hierarchical criteria of $\rm BR_{\xi}\equiv \rm BR(N_{1}\longrightarrow \xi\eta)<< \rm BR_{SM}\equiv \rm BR(N_{1}\longrightarrow SM)\simeq 1$ is considered due to the FIMP nature of $\xi$. Alongside, for the decay $N_{1}\longrightarrow \xi\eta$ to obey out of equilibrium condition, it is crucial that $\rm BR_{\xi}< 10^{-2}$. In our work, as we have solved the model parameters so as to find the value of         $ K_{N_{1}}$, we see that it is consistent in satisfying the bounds for relic abundance and streaming length. From Eq.\ref{4}, an asymptotic abundance of the FIMP DM $\xi$ can be approximated by the relation\cite{FIMP}:
				\begin{equation}
					Y_{\xi}(\infty)\simeq Y_{N_{1}}(0)\rm BR_{\xi}\big(1+\frac{15\pi\zeta(5)}{16\zeta(3)}K_{N_{1}}\big)
				\end{equation}
				Therefore, the expression for relic abundance in agreement with the asymtotic abundance is given by:
				\begin{equation}
					\Omega_{\xi}^{ FIMP}h^2= \frac{m_\xi s_{0}Y_{\xi}(\infty)}{\rho_{c}}h^2 \simeq 0.12\times \big(\frac{m_{\xi}}{keV}\big)(\rm BR_{\xi} \times 10^{3})(0.009+\frac{K_{N_{1}}}{44}))
				\end{equation}
				  where, $\rho_{c}= 1.05371 \times 10^{-5}h^{2}\rm GeV\rm cm^{-3}$ is the critical density of the Universe and $s_{0}= 2891.2\rm cm^{-3}$\cite{FIMP3} is the current entropy density and $h=0.72$ is the Hubble parameter.
				  One of the frequently and significantly considered cosmological constraints that comes into play while considering light DM candidate is the free streaming limit. It provides stringent bounds on the FIMP DM mass. Due to non- trivial velocity dispersion, the free streaming of dark matter particles strike outs the matter density perturbations and consequently the structure formation on scales $\lambda< \lambda_{FS} $. Subsequently, small structure formation gives the most robust constraint on the free streaming length, viz. $\lambda_{FS}< \mathcal{O}(0.1)$Mpc\cite{FIMP4}. The free streaming length is defined as the average distance covered by a particle without confronting collision\cite{FIMP};
				  \begin{equation}
				  	\lambda_{FS} = \int_{a_{rh}}^{a_{eq}} \frac{<v_{\xi}>}{a^{2}H}da \simeq \frac{a_{NR}}{H_{0}\sqrt{\Omega_{R}}}\big(0.62+ln\big(\frac{a_{eq}}{a_{NR}}\big)\big)
				  \end{equation}  
				  where, $<v_{\xi}>$ is the average velocity at  given time of the FIMP DM $\xi$, $a_{rh}$ and $a_{eq}$ are the scale factors at reheating and equilibrium respectively. The values of the cosmological parameters used are $H_{0}= 67.3\rm km  \rm s^{-1} \rm Mpc^{-1}$, $\Omega_{R}= 9.3\times 10^{-5}$ and $a_{eq}= 2.9\times 10^{-4}$ \cite{FIMP5}. Again, the non- relativistic scale factor for FIMP DM is expressed as\cite{lowscale}:
				  \begin{equation}
				  	a_{NR}= \frac{T_{0}}{2m_{\xi}}\big(\frac{g_{*,0}}{g_{*,rh}}\big)K_{N_{1}}^{-1/2}
				  \end{equation}  
				  with $g_{*,0}= 3.91$, $g_{*,rh}= 106.75$ and $T_{0}= 2.35\times 10^{-4}\rm eV$.
				  There are different decoupling and production mechanism of DM and thereby, the free streaming length is different for each of the relativistic decoupling scenarios such as hot, FIMP DM. As the free streaming length is highly dependent on the production mechanism, thus, the specific free streaming length for FIMP DM can be given by\cite{lowscale}:
				  \begin{equation}
				  	\lambda_{FS} \simeq 2.8\times 10^{-4}\big(\frac{keV}{m_{\xi}}\big)\big(\frac{50}{K_{N_{1}}}\big)^{1/2}\times \big(1+0.09 ln\big[\big(\frac{m_{\xi}}{keV}\big)\big(\frac{K_{N_{1}}}{50}\big)^{1/2}\big]\big)\rm Mpc
				  \end{equation} 
				  From the above equation, we can clearly say that the free streaming length is relied upon the decay parameter and the DM mass. Hence, we do a detailed analysis on these factors in our work for both NH/IH.

			\section{Analysis and results}	\label{sec6}
				\begin{figure}[h]
				\begin{center}
					\includegraphics[width=0.4\textwidth]{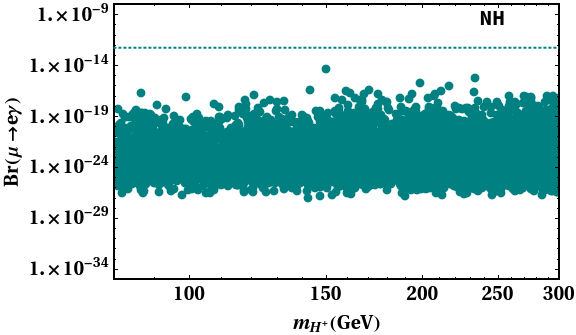}
					\includegraphics[width=0.4\textwidth]{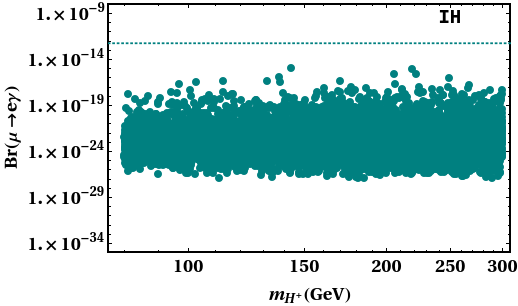}\\
					
					\caption{$Br(\mu\rightarrow e\gamma)$ as a function of $m_{H^{+}}$ for NH and IH. The dashed horizontal lines are the recent upper bounds. }\label{B1}
					
				\end{center}
			\end{figure}	
			
			\begin{figure}
				\begin{center}
					\includegraphics[width=0.4\textwidth]{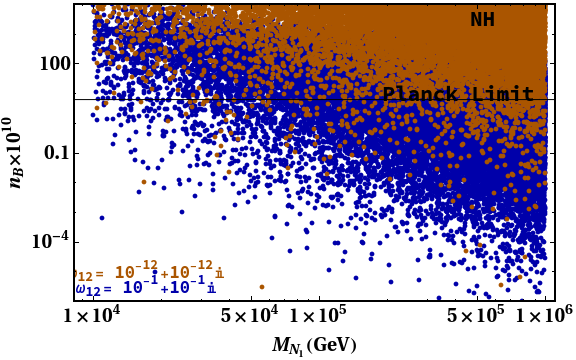}
					\includegraphics[width=0.4\textwidth]{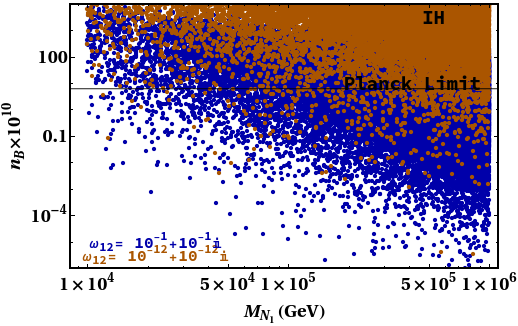}\\
					\includegraphics[width=0.4\textwidth]{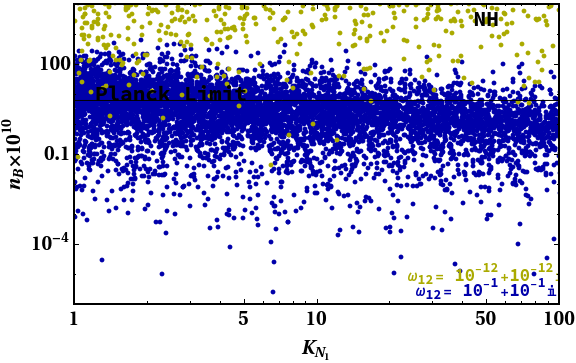}
					\includegraphics[width=0.4\textwidth]{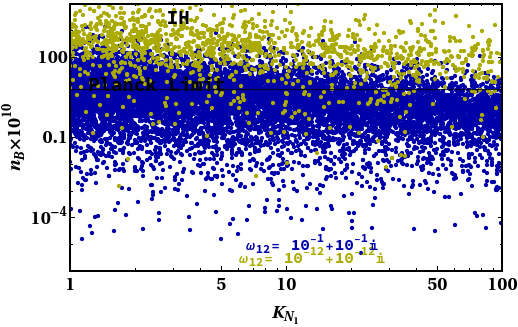}\\
					\includegraphics[width=0.4\textwidth]{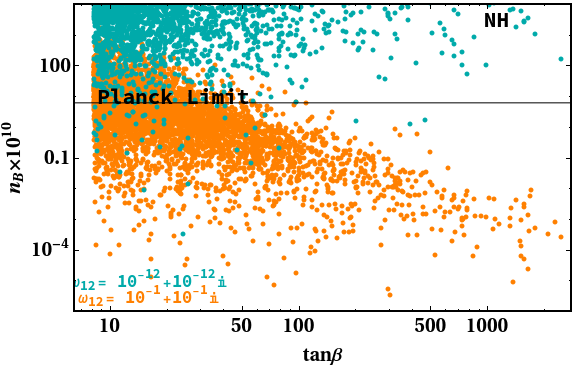}
					\includegraphics[width=0.4\textwidth]{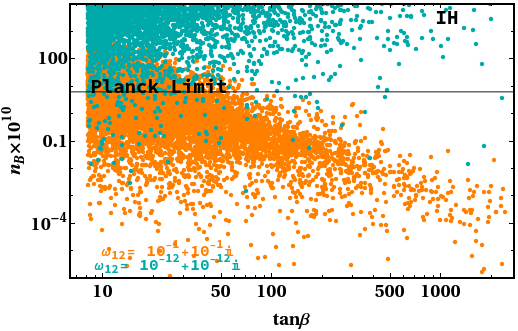}\\
				
					\caption{ Baryon asymmetry of the Universe as a function of lightest RHN($N_{1}$) is shown in the first row. The second and third row depicts BAU as a function of $ K_{N_{1}}$ and tan$\beta$ respectively. We take two different values of arbitrary angle $\omega_{12}$ and benchmark values of $\omega_{13}= 10^{-3}+10^{-3}i$ and $\omega_{23}= 10^{-2}+10^{-2}i$ for NH/IH .}\label{1}
					
				\end{center}
			\end{figure}
		
			\begin{figure}[h]
			\begin{center}
				\includegraphics[width=0.4\textwidth]{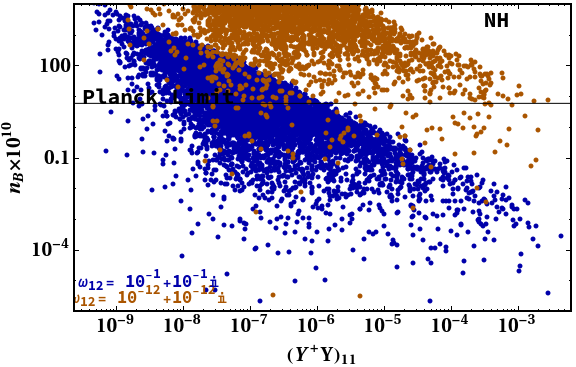}
				\includegraphics[width=0.4\textwidth]{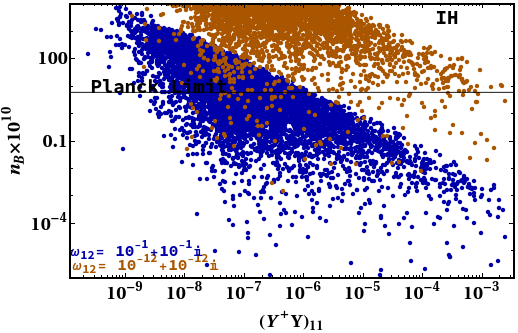}\\
				
				\caption{ Yukawa coupling term $YY^{\dagger}_{11}$ as a function of BAU for two different values of arbitrary angle $\omega_{12}$ and benchmark values of $\omega_{13}= 10^{-3}+10^{-3}i$ and $\omega_{23}= 10^{-2}+10^{-2}i$ for NH/IH . }\label{21}
				
			\end{center}
		\end{figure}	
		
		\begin{figure}[h]
			\begin{center}
				\includegraphics[width=0.4\textwidth]{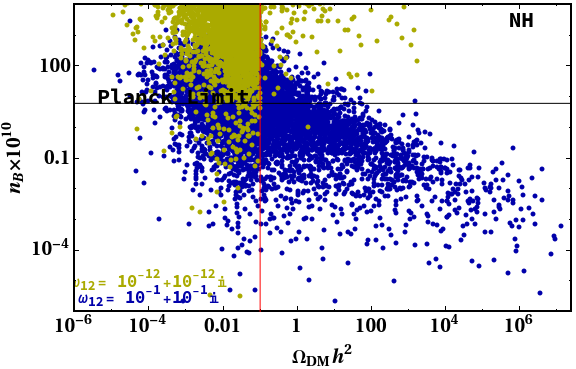}
				\includegraphics[width=0.4\textwidth]{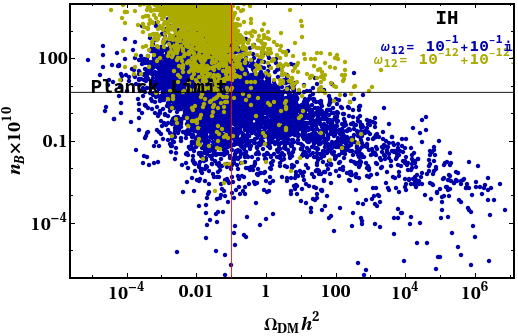}\\
				
				\caption{ Baryon asymmetry of the Universe as a function of relic abundance of the dark matter candidate $\xi$ for NH/IH with the consideration of two different values of arbitrary angle $\omega_{12}$ and benchmark values of $\omega_{13}= 10^{-3}+10^{-3}i$ and $\omega_{23}= 10^{-2}+10^{-2}i$. }\label{12}
				
			\end{center}
		\end{figure}
	
		\begin{figure}[h]
		\begin{center}
			\includegraphics[width=0.4\textwidth]{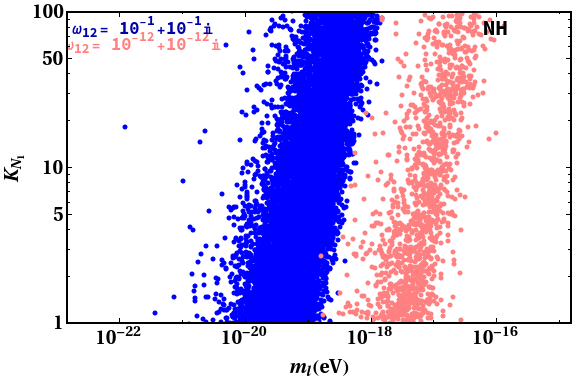}
			\includegraphics[width=0.4\textwidth]{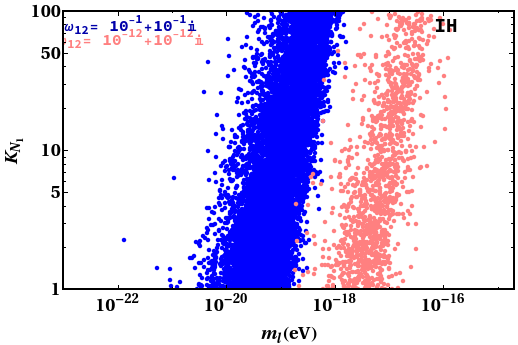}\\
			
			\caption{ Decay parameter $ K_{N_{1}}$ as a function of lightest neutrino mass for two different values of arbitrary angle $\omega_{12}$ with benchmark values $\omega_{13}= 10^{-3}+10^{-3}i$ and $\omega_{23}= 10^{-2}+10^{-2}i$ for NH/IH . }\label{14}
			
		\end{center}
	\end{figure}	

	\begin{figure}[h]
	\begin{center}
		\includegraphics[width=0.4\textwidth]{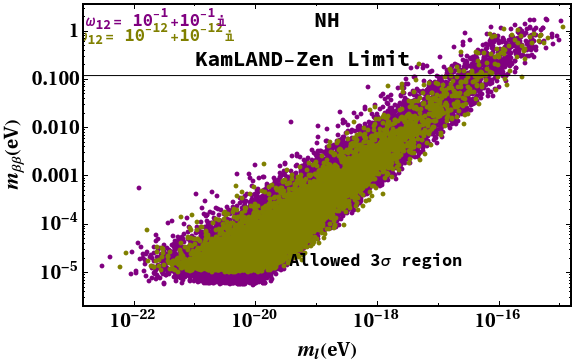}
		\includegraphics[width=0.4\textwidth]{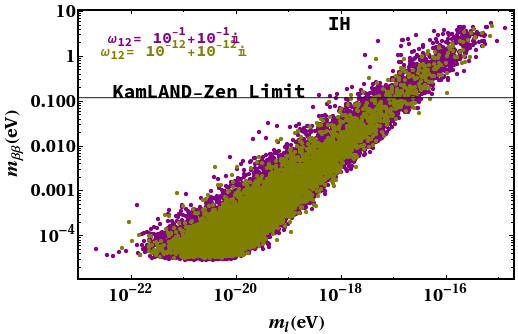}\\
		
		\caption{ Effective mass of the neutrinos w.r.t the lightest active neutrino is being showed for NH/IH with two different values of arbitrary angle $\omega_{12}$ and benchmark values of $\omega_{13}= 10^{-3}+10^{-3}i$ and $\omega_{23}= 10^{-2}+10^{-2}i$ . The horizontal black line signify the KamLAND-Zen upper limit on the effective mass of neutrinos. }\label{15}
		
	\end{center}
\end{figure}		
	\begin{figure}[h]
	\begin{center}
		\includegraphics[width=0.4\textwidth]{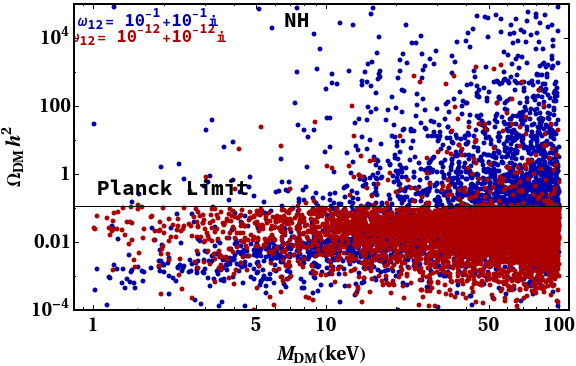}
		\includegraphics[width=0.4\textwidth]{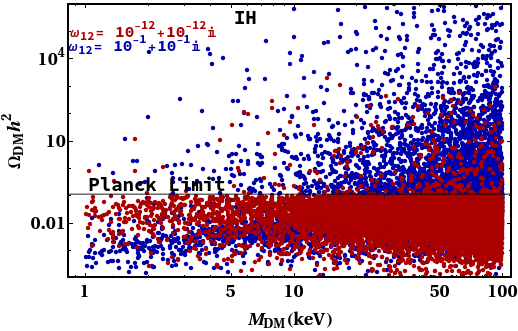}\\
		
		\caption{The plot shows variation of relic abundance w.r.t the dark matter mass for NH/IH. Two different values of arbitrary angle $\omega_{12}$ and benchmark values of $\omega_{13}= 10^{-3}+10^{-3}i$ and $\omega_{23}= 10^{-2}+10^{-2}i$ for NH/IH is chosen . }\label{16}
		
	\end{center}
\end{figure}		
		
			\begin{figure}[h]
			\begin{center}
				\includegraphics[width=0.4\textwidth]{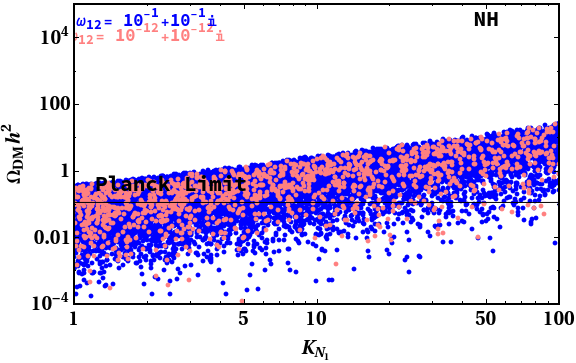}
				\includegraphics[width=0.4\textwidth]{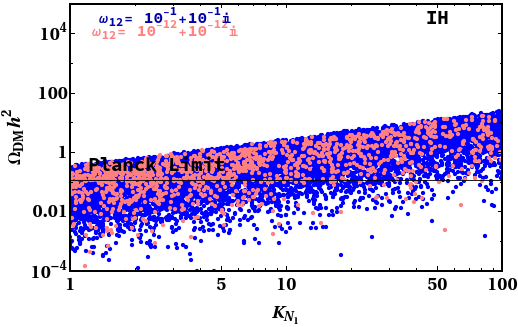}\\
				
				\caption{The plot shows variation of relic abundance w.r.t the decay parameter $ K_{N_{1}}$ for NH/IH for two different values of arbitrary angle $\omega_{12}$ and benchmark values of $\omega_{13}= 10^{-3}+10^{-3}i$ and $\omega_{23}= 10^{-2}+10^{-2}i$ . }\label{17}
				
			\end{center}
		\end{figure}	
	
		\begin{figure}[h]
		\begin{center}
			\includegraphics[width=0.4\textwidth]{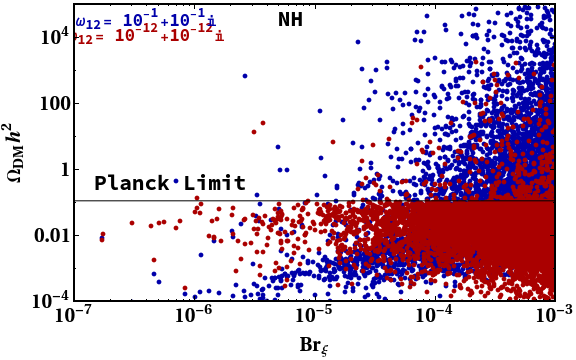}
			\includegraphics[width=0.4\textwidth]{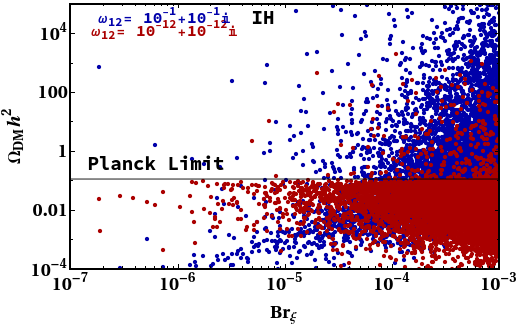}\\
			
			\caption{The plot shows variation of relic abundance w.r.t the branching ratio of the decay $N_{1}\longrightarrow \xi \chi$ for two different values of arbitrary angle $\omega_{12}$ and benchmark values of $\omega_{13}= 10^{-3}+10^{-3}i$ and $\omega_{23}= 10^{-2}+10^{-2}i$ for NH/IH. }\label{18}
			
		\end{center}
	\end{figure}	

	\begin{figure}[h]
	\begin{center}
		\includegraphics[width=0.4\textwidth]{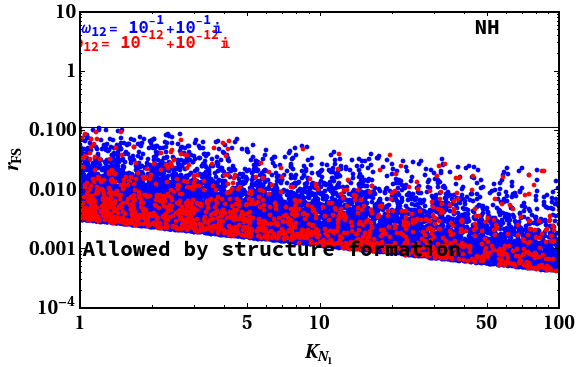}
		\includegraphics[width=0.4\textwidth]{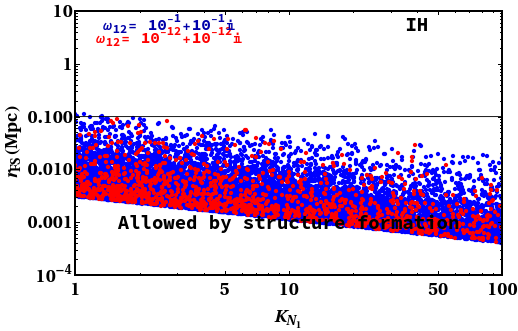}\\
		
		\caption{The plot shows variation of free streaming length w.r.t the decay parameter $ K_{N_{1}}$ for NH/IH. Here, the variation is shown for two different values of arbitrary angle $\omega_{12}$ and benchmark values of $\omega_{13}= 10^{-3}+10^{-3}i$ and $\omega_{23}= 10^{-2}+10^{-2}i$. }\label{19}
		
	\end{center}
\end{figure}	

	\begin{figure}[h]
	\begin{center}
		\includegraphics[width=0.4\textwidth]{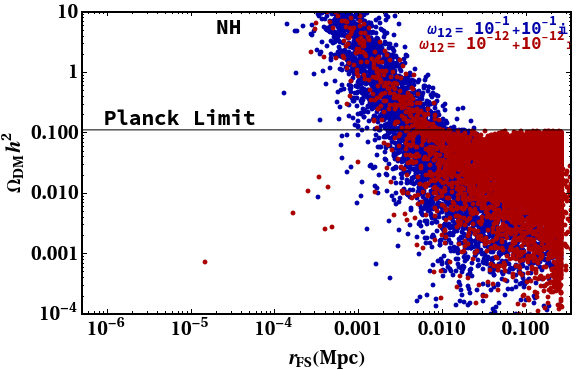}
		\includegraphics[width=0.4\textwidth]{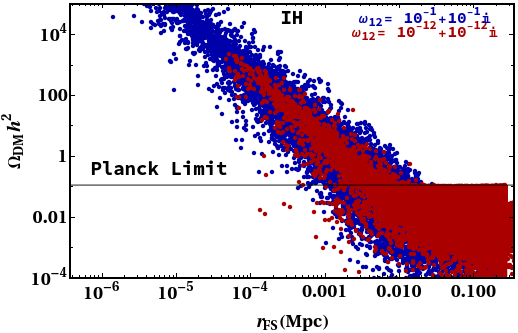}\\
		
		\caption{The plot shows variation of free streaming length w.r.t the relic abundance of $\xi$ for two different values of arbitrary angle $\omega_{12}$ and benchmark values of $\omega_{13}= 10^{-3}+10^{-3}i$ and $\omega_{23}= 10^{-2}+10^{-2}i$ for NH/IH. }\label{20}
		
	\end{center}
\end{figure}

\begin{figure}
	\begin{center}
		\includegraphics[width=0.4\textwidth]{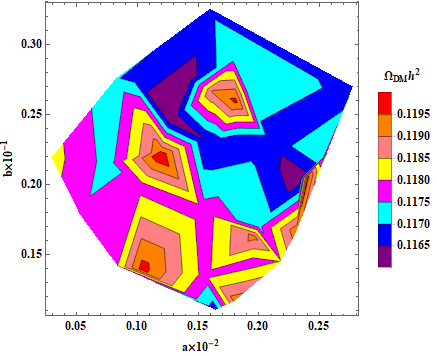}
		\includegraphics[width=0.4\textwidth]{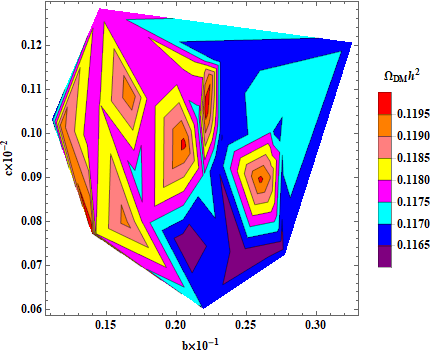}\\
		\includegraphics[width=0.4\textwidth]{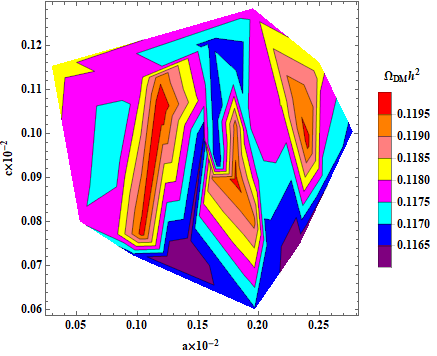}
		
		\caption{Contour plot relating the model parameters a and b w.r.t the relic abundance($\Omega_{DM}h^{2}$) of the DM candidate($\xi$)(left panel), right panel shows variation of b and c with $\Omega_{DM}h^{2}$ and in the middle we show a similar plot for a and c. This result corresponds to benchmark points $\omega_{12}= 10^{-1}+10^{-1}i$, $\omega_{13}= 10^{-3}+10^{-3}i$ and $\omega_{23}= 10^{-2}+10^{-2}i$ for NH.  }\label{c1}
			\end{center}
	\end{figure}
\begin{figure}
	\begin{center}
		\includegraphics[width=0.4\textwidth]{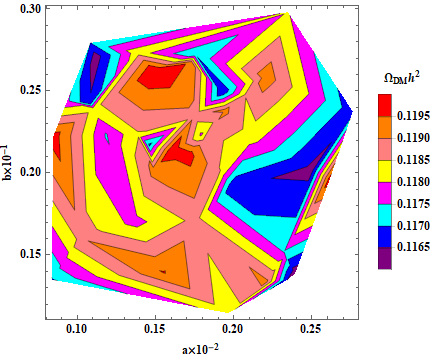}
		\includegraphics[width=0.4\textwidth]{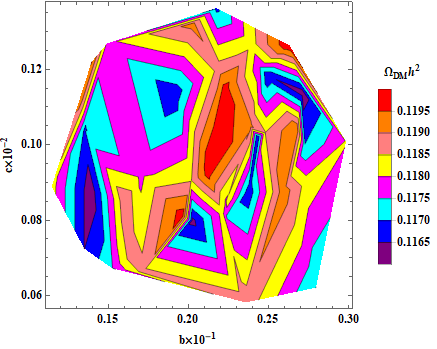}\\
		\includegraphics[width=0.4\textwidth]{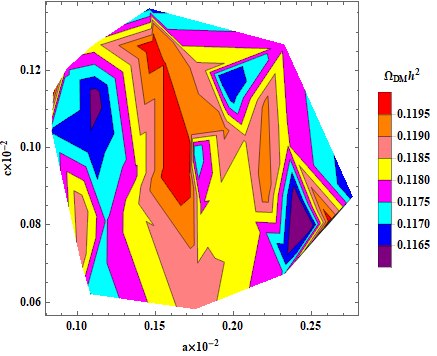}
		
		\caption{Contour plot relating the model parameters a and b w.r.t the relic abundance($\Omega_{DM}h^{2}$) of the DM candidate($\xi$)(left panel), right panel shows variation of b and c with $\Omega_{DM}h^{2}$ and in the middle we show a similar plot for a and c. This result corresponds to benchmark points $\omega_{12}= 10^{-1}+10^{-1}i$, $\omega_{13}= 10^{-3}+10^{-3}i$ and $\omega_{23}= 10^{-2}+10^{-2}i$ for IH.  }\label{c2}
	\end{center}
\end{figure}

\begin{figure}
	\begin{center}
		\includegraphics[width=0.4\textwidth]{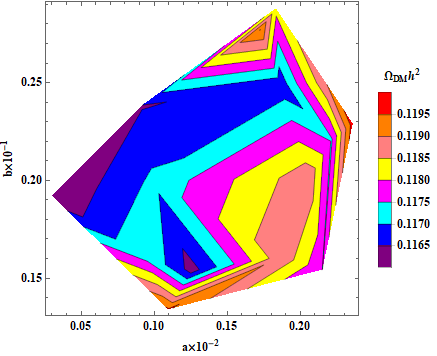}
		\includegraphics[width=0.4\textwidth]{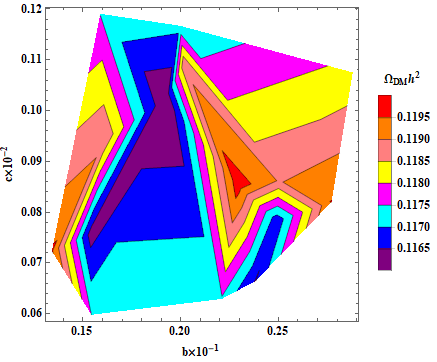}\\
		\includegraphics[width=0.4\textwidth]{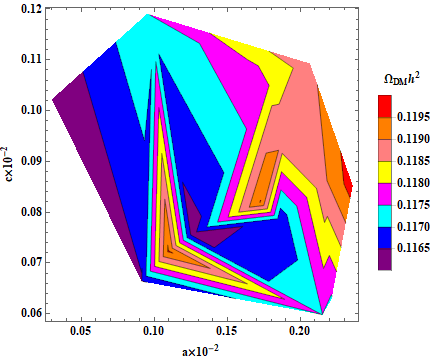}
		
		\caption{Contour plot relating the Dirac masses a and b w.r.t the relic abundance($\Omega_{DM}h^{2}$) of the DM candidate($\xi$)(left panel), right panel shows variation of b and c with $\Omega_{DM}h^{2}$ and in the middle we show a similar plot for a and c. This result corresponds to benchmark points $\omega_{12}= 10^{-12}+10^{-12}i$, $\omega_{13}= 10^{-3}+10^{-3}i$ and $\omega_{23}= 10^{-2}+10^{-2}i$ for NH.  }\label{c3}
	\end{center}
\end{figure}

 \begin{figure}
 	\begin{center}
 		\includegraphics[width=0.4\textwidth]{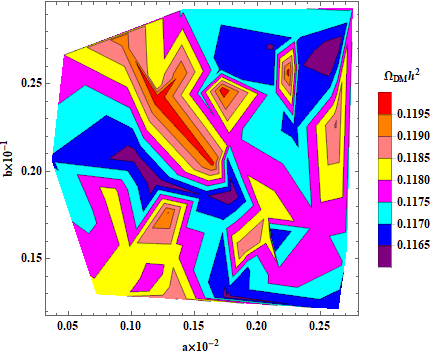}
 		\includegraphics[width=0.4\textwidth]{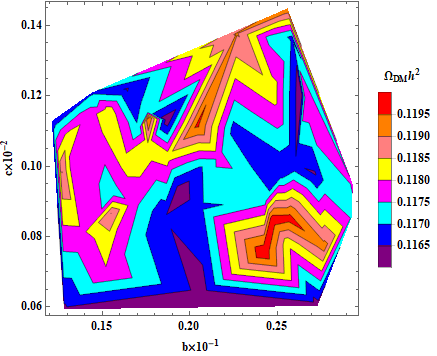}\\
 		\includegraphics[width=0.4\textwidth]{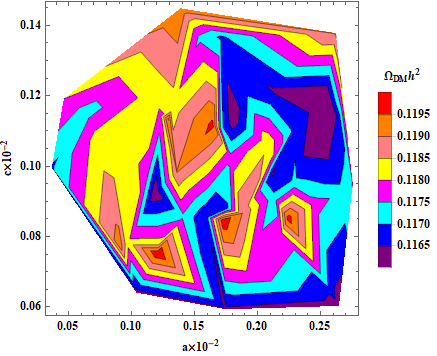}
 		
 		\caption{Contour plot relating the Dirac masses a and b w.r.t the relic abundance($\Omega_{DM}h^{2}$) of the DM candidate($\xi$)(left panel), right panel shows variation of b and c with $\Omega_{DM}h^{2}$ and in the middle we show a similar plot for a and c. This result corresponds to benchmark points $\omega_{12}= 10^{-12}+10^{-12}i$, $\omega_{13}= 10^{-3}+10^{-3}i$ and $\omega_{23}= 10^{-2}+10^{-2}i$ for IH.  }\label{c4}
 	\end{center}
 \end{figure}

In the flavor symmetric realisation of the extended $\nu$2HDM\cite{lowscale}, we have explicitly studied the implication of the Yukawa coupling matrix and the rotational matrix used for parametrization. On numerically solving the Dirac masses given by a,b and c which constitute the Yukawa coupling matrix, we eventually find the range of the coupling matrix Y. Incorporating the parameter space of the Yukawa coupling matrix in Eq.\eqref{yuk}, we can deduce the active neutrino mass matrix. It has been found from the model we have worked upon that the summation of the active neutrino is consistent with the Planck limit, i.e. $\sum_{i=1,2,3}m_{i} < 0.11 eV$\cite{Lattanzi:2016rre}. Consecutively, we have also analysed the phenomenology of $0\nu\beta\beta$\cite{Mohapatra:1986su,Barry:2013xxa,Borgohain:2017inp} in our work. It is indeed a well known and significant experimental technique of detecting neutrino mass, thereby, making it relatable with observations in the on-going experiments. The various experiments in this regard are KamLAND-Zen\cite{Kamland2,kamland}, KATRIN\cite{katrin2,KATRIN}, GERDA\cite{gerda,GERDA2}, etc. The phenomenon that is measured in these experiments is the effective mass of the active neutrinos which is given by the expression:
\begin{equation}
|m_{\beta\beta}|= |c_{12}^{2}c_{13}^{2}m_{1} + s_{12}^{2}c_{13}^{2}m_{2}e^{2i\alpha} + s_{13}^{2}m_{3}e^{2i\beta}|
\end{equation}
where, $c_{ij}$= $\cos\theta_{ij}$ and $s_{ij}$= $\sin\theta_{ij}$ are the elements of the UPMNS matrix. It is crucial to obey the effective mass constraint so as to make the model sensitive to the on-going as well as future collider signatures.   
Throughout the work, we have considered a specific range of the right handed neutrinos, viz. $M_{N_{1}}= 10^4$ GeV - $10^{6}$ GeV, $M_{N_{2}}= 10^7$ GeV - $5\times 10^{7}$ GeV and $M_{N_{3}}= 10^8$ GeV - $5\times 10^{8}$ GeV. Also, the lightest active neutrino mass is taken in the span of $m_{l}= 10^{-11}- 10^{-13}$ eV in case of both NH/IH which is a crucial parameter required to achieve baryogenesis in TeV scale\cite{Hugle:2018qbw}. In our study, we have shown the variations that may occur due to the choice of arbitrary angles of the rotational matrix given in Eq.\eqref{2}. Since, the dependency of BAU and dark matter phenomenology is influenced by the decay parameter $ K_{N_{1}}$, thus we can generate desired values of $ K_{N_{1}}$ by fine tuning the arbitrary angles $\omega_{12}$ and $\omega_{13}$ \cite{lowscale}. In our case, we have kept $\omega_{13,23}$ fixed at $10^{-3}+10^{-3}i$	and $10^{-2}+10^{-2}i$ respectively and simultaneously chosen two different values of $\omega_{12}$ as $10^{-1}+10^{-1}i$ and $10^{-12}+10^{-12}i$. In Fig.\eqref{1}, we have shown co-relation plots between BAU and parameters such as $M_{N_{1}}$, $ K_{N_{1}}$ and tan$\beta$ respectively for NH and IH. From the first row of Fig.\eqref{1}, we see that the observed baryogenesis is generated for $M_{N_{1}}=  10^{4}- 10^{6}$ GeV for NH, however, the choice of the arbitrary angle $\omega_{12}= 10^{-1}+10^{-1}i$ gives large number of points satisfying the Planck limit compared to that of $\omega_{12}= 10^{-12}+10^{-12}i$. For $\omega_{12}= 10^{-12}+10^{-12}i$, the region $M_{N_{1}}= 5\times 10^{4}- 10^{6}$GeV has prominent points satisfying the desired BAU in case of NH. For IH, we donot observe much change in the variation of BAU w.r.t $M_{N_{1}}$ obeying the Planck bound. Also the variation due to the arbitrary angles are almost similar for both the heirarchies. In the second row of Fig.\eqref{1}, an interesting result is seen which constraints the parameter space of $ K_{N_{1}}$ satisfying the Planck limit depending upon the choice of arbitrary angle $\omega_{12}$. The points satisfying BAU for $\omega_{12}= 10^{-12}+10^{-12}i$ are very scanty, whereas for $\omega_{12}= 10^{-1}+10^{-1}i$ it is seen that we have abundant points for the decay parameter space $K_{N_{1}}= 10-100$, thereby producing thermal leptogenesis and also serving the DM candidate to be a FIMP type for both NH/IH \cite{lowscale,FIMP}. Again from third row of Fig.\eqref{1}, we have obtained a clear distinction between the range of tan$\beta$ satisfying the Planck limit for BAU w.r.t the choice of $\omega_{12}$. We see that for $\omega_{12}= 10^{-1}+10^{-1}i$, the range of tan$\beta$ from 10-100 is seen to produce the observed BAU, whereas for $\omega_{12}= 10^{-12}+10^{-12}i$ the region of tan$\beta$ satisfying the BAU limit is constrained from 10-50 with very less points as compared to that for $\omega_{12}= 10^{-1}+10^{-1}i$ incase of NH. The results obtained for IH is almost similar to that for NH. We have obtained a crucial result from Fig.\eqref{21} which remains almost the same for both the mass heirarchies. Here, the parameter space of the Yukawa coupling element $(Y^{\dagger}Y)_{11}$ which has a significant contribution in the CP asymmetry is constrained for $\omega_{12}= 10^{-12}+10^{-12}i$ in a small region obeying the BAU limit. Also we have obtained very few points in the parameter space satisfying the Planck bound incase of lower value of $\omega_{12}$. However, we have quite a wide range of $(Y^{\dagger}Y)_{11}$ corresponding to $\omega_{12}= 10^{-1}+10^{-1}i$ satisfying the Planck bound for BAU, i.e. $(Y^{\dagger}Y)_{11}= 10^{-8}-10^{-5}$ . Fig.\eqref{12} is a co-relation plot between the relic abundance and the BAU. Here, we see that there are sufficient common points for the choice of $\omega_{12}=10^{-1}+10^{-1}i$ satisfying both the bounds simultaneously, though we have almost negligible number of points obeying both the bounds incase of $\omega_{12}=10^{-12}+10^{-12}i$. The results are similar for both the heirarchies.

 \begin{table}
	\begin{center}
		\begin{tabular}{|c|c|c|}
			
			\hline 
			Parameter&NH&IH  \\ 
			\hline
			a(GeV)&$0.05\times10^{-2}-0.28\times10^{-2}$&$0.10\times10^{-2}-0.25\times10^{-2}$  \\
			\hline 
				b(GeV)&$0.12\times10^{-1}-0.31\times10^{-1}$&$0.12\times10^{-1}-0.30\times10^{-1}$ \\
			\hline  
				c(GeV)&$0.06\times10^{-2}-0.13\times10^{-2}$&$0.06\times10^{-2}-0.14\times10^{-2}$ \\
			\hline 	
		\end{tabular} 
	\end{center}
	\caption{Dirac masses of the model and their respective parameter space satisfying the relic abundance of the DM candidate $\xi$ for $\omega_{12}=10^{-1}+10^{-1}i$.} \label{TAB3}
\end{table}

\begin{table}
	\begin{center}
		\begin{tabular}{|c|c|c|}
			
			\hline 
			Parameter&NH&IH  \\ 
			\hline
				a(GeV)&$0.05\times10^{-2}-0.24\times10^{-2}$&$0.05\times10^{-2}-0.25\times10^{-2}$  \\
			\hline 
				b(GeV)&$0.14\times10^{-1}-0.28\times10^{-1}$&$0.13\times10^{-1}-0.28\times10^{-1}$ \\
			\hline  
				c(GeV)&$0.06\times10^{-2}-0.12\times10^{-2}$&$0.06\times10^{-2}-0.14\times10^{-2}$ \\
			\hline 	
		\end{tabular} 
	\end{center}
	\caption{Dirac masses of the model and their respective parameter space satisfying the relic abundance of the DM candidate $\xi$ for $\omega_{12}=10^{-12}+10^{-12}i$.} \label{TAB4}
\end{table}

From Fig.\eqref{14}, we see a divergence in the plot of decay parameter w.r.t the lightest active neutrino mass. For $\omega_{12}= 10^{-1}+10^{-1}i$, the decay parameter $ K_{N_{1}}$ in the allowed region, i.e. 10-100 from the thermal leptogenesis point of view is obtained for light neutrino mass in the space $10^{-19}-10^{-21}$. This range of $K_{N_{1}}$ corresponds to warm DM samples. Whereas, for  $\omega_{12}= 10^{-12}+10^{-12}i$, there is a shift in the mass range of lightest active neutrino. 
 However, there is hardly any change in the effective mass curve depicted in Fig.\eqref{15} irrespective of the two different choice of $\omega_{12}$ for NH as well as IH. Typically for a warm DM, the branching ratio $\rm Br_{\xi}> 10^{-4}$, whereas for a cold DM it is $\rm Br_{\xi}\lesssim 10^{-3}$. For further deduction of the relic abundance numerically, we choose the branching ratio in the range $\rm Br_{\xi}= 10^{-2}- 10^{-7}$. From Fig.\eqref{16}, we see that for $\omega_{12}= 10^{-1}+10^{-1}i$, we get observed relic abundance for DM mass in the range 5-100 GeV which corresponds to warm DM, whereas for $\omega_{12}= 10^{-12}+10^{-12}i$, smaller mass of $\xi$ signifying hot DM also satisfies the relic abundance limit. In a plot between relic abundance and decay parameter($ K_{N_{1}}$) as shown in Fig.\eqref{17}, we observe that the parameter space for $ K_{N_{1}}=10-100$ satisfies the Planck limit for relic abundance for $\omega_{12}= 10^{-1}+10^{-1}i$ whereas for $\omega_{12}= 10^{-1}+10^{-1}i$, the same region of $ K_{N_{1}}$ hardly satisfies relic abundance constraint. However, for satifying the small structure formation constraint, $ K_{N_{1}}$ must not fall in the very weak washout region, therefore, $\omega_{12}= 10^{-12}+10^{-12}i$ is less favorable for both NH/IH. Also as mentioned earlier the preferable branching ratio range for warm DM satisfying the relic abundance bound is obeyed for $\omega_{12}= 10^{-1}+10^{-1}i$ as depicted in Fig.\eqref{18}. We know that the most stringent bound on $r_{FS}$ comes from the small structure formation, $r_{FS}< 0.1$. A relationship between $r_{FS}$ and $ K_{N_{1}}$ is shown in Fig.\eqref{19}, where we have all the points for both the cases of $\omega_{12}$ in the allowed region. As we know that $ K_{N_{1}}$ is inversely proportional to $r_{FS}$, thus, very small values of $ K_{N_{1}}$ donot satisfy the small structure formation bound. An analysis on the points satisfying both relic abundance and small structure formation is been shown in Fig.\eqref{20}, wherein we see that almost all the points corresponding $r_{FS}= 0.01-0.1$ satisfy the relic abundance bound. Also from Fig.\eqref{16}, we have obtained the allowed range of warm DM mass satisying the relic abundance to be 5-100 KeV, thus we can summarise that this range of DM mass also abides the streaming length constraint. We have further shown some contour plots between the Dirac masses a, b and c w.r.t the allowed range for relic abundance of the dark matter, i.e. $\Omega_{DM}h^{2}= 0.1186 \pm 0.0020$ given by Planck data. Thus, in Fig.\ref{c1},\ref{c2} we have shown the parameter space of a, b and c which satisfies the relic abundance limit corresponding to $\omega_{12}=10^{-1}+10^{-1}i$ for both NH and IH respectively. Again similar plots are depicted for $\omega_{12}=10^{-12}+10^{-12}i$ in Fig.\ref{c3},\ref{c4}. Summarising the allowed parameter space of the Dirac masses from the above mentioned figures, we finally represent it in tabular form in Table.\ref{TAB3} and Table.\ref{TAB4}.

\section{Conclusion}\label{sec7}
Our work basically showcases a comparative analysis of different phenomenological consequences corresponding to variation in the arbitrary angles of the rotational matrix. As mentioned in the earlier sections, we have realised the extended $\nu$2HDM with the help of flavor symmetries $A_{4}\otimes Z_{4}$. It is an interesting model as it can accomodate both neutrino as well as DM phenomenologies. Also, we can generate the light neutrino mass at the tree level by consideration of an extra scalar doublet which is assigned L=-1\cite{nu2HDM}. From the model, we numerically calculate the Yukawa coupling matrix, which plays the source for the various phenomenologies that we have studied. Also, we have chosen the rotational matrix R and the values of the arbitrary angles such that we can draw crucial conclusions of its impact on the cosmological phenomena we have discussed. Following various literatures\cite{Hugle:2018qbw,FIMP5}, we have chosen the values of the RHN masses, lightest active neutrino mass in the range $10^{-13}-10^{-11}$eV, vev of the scalar doublet($\phi$) as already mentioned in sec.\eqref{sec6} so as to achieve TeV scale leptogenesis. We have also co-related the BAU with the values of tan$\beta$ and analyzed its variation for different $\omega_{12}$ values. From Fig.\eqref{1},\eqref{21} and \eqref{12} we can conclude that the choice $\omega_{12}= 10^{-1}+10^{-1}i$ is more preferable in satisfying the BAU limit given by Planck data. Also, we have obtained a constraint range of the free parameters as well as the Yukawa couplings corresponding to the two choices of the arbitrary angles. In plot \eqref{12} we obtain points which abide by the constraint from both BAU and relic abundance simultaneously which is a vital part of the analysis. We have also generated some important results in context with the decay parameter. Depending on $\omega_{12}$ values, we can generalise the transition in the relation between $ K_{N_{1}}$ and $m_{l}$ for both NH and IH. Considering the DM scenario, it is viable to have a warm DM candidate as it obeys the constraint coming from small structure formation and relic abundance  w.r.t to the allowed parameter space of decay parameter in comparison to hot DM. Simultaneously, we can get a warm DM source when the decay parameter donot fall in the very weak washout region. Therefore, it is seen from Fig.\eqref{14} that for $\omega_{12}=10^{-1}+10^{-1}i$ we obtain the preferable range of $ K_{N_{1}}$ which further explains the warm DM. For warm DM, $m_{\xi}\sim 10$ KeV gives significant observed relic abundance, however, small DM mass correspond to hot DM which again is constrained by small structure formation. Again, a difference in the branching ratio range is obtained depending on the two different values of $\omega_{12}$. So, we can say that $\rm Br_{\xi}\simeq 10^{-5}-10^{-3}$ is the prefered range for $\omega_{12}= 10^{-1}+10^{-1}i$ obeying the Planck limit for relic abundance, whereas for $\omega_{12}= 10^{-12}+10^{-12}i$, $\rm Br_{\xi}\simeq 10^{-6}-10^{-3}$ produces the observed relic abundance. Summarizing the above results, it can be said that for the choice of $\omega_{12}= 10^{-1}+10^{-1}i$, the value of $ K_{N_{1}}$ falls in the weak washout region which further is successful in generating the desired BAU, relic abundance and also the small structure formation for both NH and IH. Also, the Yukawa couplings obtained from the model are successful in producing the branching ratio Br$(\mu \rightarrow e\gamma)$$<4.2 \times 10^{-13}$ as shown in Fig.\eqref{B1}. Thus, the model can explain the neutrino mass, leptogenesis and dark matter for the choice of the free parameters considered and the values of Dirac masses obtained.  

\section{Acknowledgement}
The research work of MKD is supported by the Department of Science and Technology, Government of India, under the project grant EMR/2017/001436. LS and BBB would like to acknowledge Tezpur University institutional grant and Research and Innovation grant DoRD/RIG/10-73/ 1592-A for funding their research work. 
\bibliographystyle{utphys}
\bibliography{ref1}
		\end{document}